# SN 1986E eight years after outburst: a link to SN 1957D? *

E. Cappellaro,[1], I. J. Danziger[2], M. Turatto[1]
[1] *Osservatorio Astronomico di Padova, vicolo dell'Osservatorio 5, I-35122 Padova, Italy*
[2] *European Southern Observatory, Karl-Schwarzschild-Strasse 2, D-85748 Garching bei München, Germany*

**ABSTRACT**

We report on the optical recovery of SN 1986E 7.1 and 7.9 years after explosion. The spectra are dominated by broad H$\alpha$ and [O I] $\lambda\lambda 6300,6364$ emission lines with a minor contribution from [Ca II] $\lambda\lambda 7291,7324$ and, possibly, [Fe II] $\lambda 7155$. Compared with the spectra of the other recovered SNII, SN1986E shows strong H$\alpha$ luminosity similar to SN 1979C but, in contrast, weak [O III] emission. So far, among classified SNe, only type II Linears have been recovered a few years past the explosion. The high optical luminosity of Linear SNII at this phase is most likely related to the interaction of the ejecta with the circumstellar wind which is also expected to produce strong radio emission. Therefore radio observations of the remnant of SN 1986E are worth attempting.

The circumstellar interaction model predicts that, in the next decade, the strength of the [O III]/H$\alpha$ emission rapidly increases and [O III] $\lambda\lambda 4959,5007$ should eventually dominate the spectrum. We observed this effect in SN 1957D 30 years after explosion which therefore may represent the normal evolution of SNII with a dense circumstellar medium. The details observed in the spectra of both SNe suggest directions for further refinement of the modelling.

## 1 INTRODUCTION

With the exception of the nearby SN 1987A, SNe usually fade below the detection limit of even the largest telescope in just a couple of years. However, thanks the development of the instrumentation, in the last few years a handful of SNe have been optically recovered at ages from 5 to 35 yr: of these SN 1957D, SN 1978K and SN 1986J were not observed near maximum, SN 1961V was a peculiar, presently unique object, SN 1988Z had from discovery an unusually slow luminosity decline, and all the other three, SN 1970G, SN 1979C and SN 1980K, were classified type II Linear (Leibundgut 1991; Fesen 1993). Even from these poor statistics, there is the suggestion that, at these phases, the luminosity of Linear SNII is higher than that of Plateau SNII, SNIa or SNIb/c.

The subdivision of SNII into Plateau and Linear, proposed by Barbon et al. (1979), is based on the photometric evolution in the first three months after explosion. There are a number of intermediate cases in which an almost linear decline is interrupted by a short plateau phase: SN 1970G was an example of this kind so that, in the literature, it has been classified alternatively as Linear and Plateau (Barbon et al. 1979; Turatto et al. 1990).

A quantitative criterion for the classification of SNII light curves was proposed by Patat et al. (1994) based on the rate of luminosity decline in the early 100 days ($\beta_{100}$): Linear SNII have a decline rate $\beta^B_{100} > 3.5$ mag 100d$^{-1}$ (according to this criterion SN 1970G having $\beta^B_{100} = 4.7$ mag 100d$^{-1}$ was classified Linear). Starting 150 days after explosion, the luminosities of both Plateau and Linear SNII settle onto a slower exponential decline, $\gamma^V \sim 0.9$ mag 100d$^{-1}$ (Turatto et al. 1990), consistent with the radioactive decay of $^{56}$Co into $^{56}$Fe. There are not strong spectral differences between the two SNII types but, in general, Linear SNII have shallower P-Cygni troughs than Plateau SNII (Patat et al. 1994) which is indicative of different envelope structures/masses.

SN 1986E was discovered on April 13, 1986 (*J.D.* = 2446534) by G. Candeo at the Asiago Observatory at a magnitude $B = 14.2$. The parent galaxy NGC 4302 is an edge-on *Sc* and the SN appeared, in projection, about 1 kpc above the galactic plane and very close (7.0″E and 0.6″S) to a 15 mag foreground star. Observations were obtained at Asiago shortly after the discovery but, although the SN was relatively bright, only incomplete coverage was obtained (Cappellaro et al. 1989). It was classified as type II because of the presence of H$\alpha$ in the early spectrum. The shape of the light curve and in particular the decline rate in the first 100 days, $\beta_{100}^{B} = 4.8$ mag 100d$^{-1}$, is typical of the Linear SNII.

---

* Based on observations collected at ESO - La Silla (Chile).

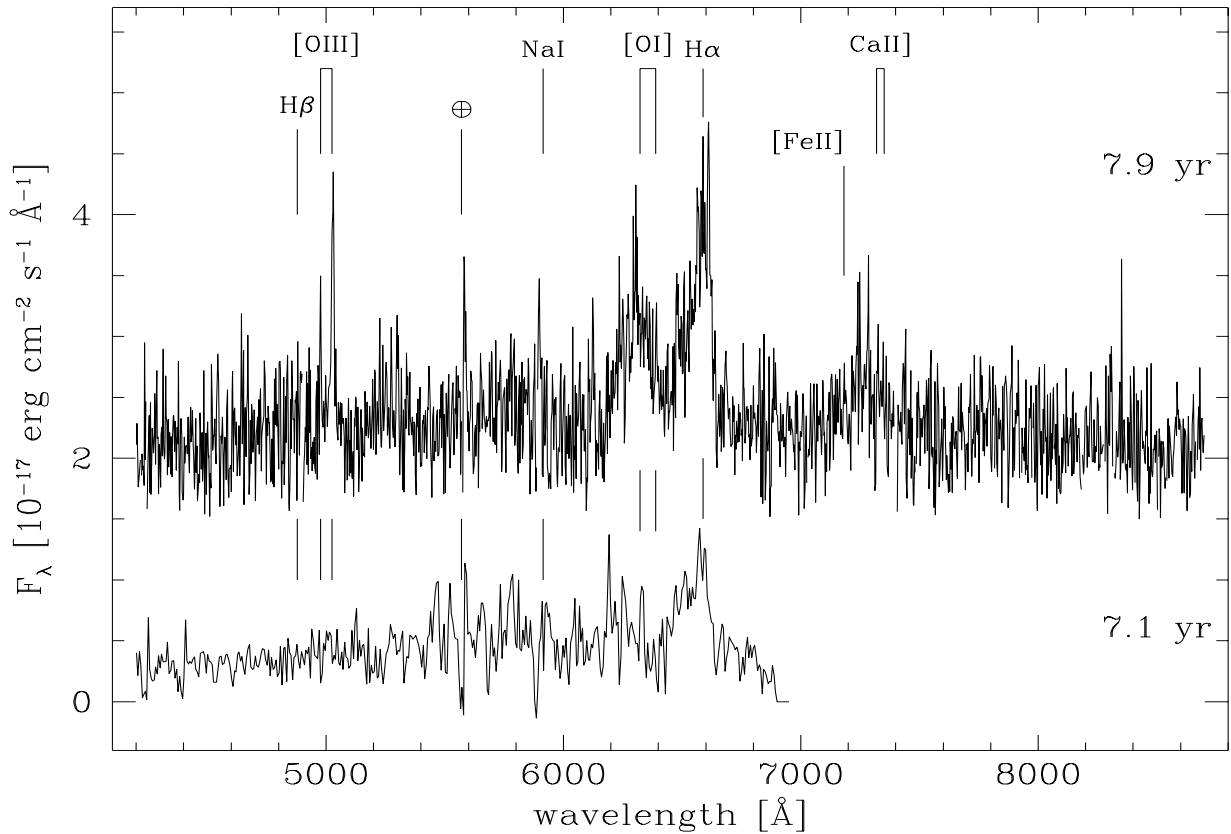

**Figure 1.** Spectra of SN1986E 7.1 yr (3.6m + EFOSC) and 7.9 yr (NTT + EMMI) after explosion. The latter has been shifted in the vertical axis by 2 units. Vertical lines mark the positions of the lines in the galaxy rest frame.

Only sporadic observations were obtained during the radioactive tail period (phase > 150 d). Spectra obtained by Filippenko and Shields (1986) and by Schlegel et al. (1995) respectively 9 and 10 months after discovery revealed strong H$\alpha$, [O I] $\lambda\lambda$6300,6364 and [Ca II] $\lambda\lambda$7292,7324 lines, typical of SNII at this phase (Cappellaro et al. 1994; Clocchiatti et al. 1995). In addition a photometric measurement 660 days after maximum gave $R = 21.5$ for the SN magnitude, fully consistent with a normal exponential tail (Turatto et al. 1990).

In the following we present the optical spectra plus some photometric points of SN 1986E obtained at ESO-La Silla 7.1 and 7.9 yr past maximum and we compared them with the ESO observations of SN 1957D 31.3 and 33.1 yr after outburst. All observations were obtained within the framework of the ESO Key Programme on SNe.

## 2 OBSERVATIONS

Observations of SN 1957D were obtained using the ESO 3.6m telescope equipped with the ESO Faint Object Spectrograph and Camera (EFOSC), on April 7-8, 1989 (31.3 yr, cfr. Turatto et al. 1989) and on February 21, 1991 (33.1 yr). Both B300 and R300 grisms were used that, in combination with the RCA CCD #8, allow the coverage of the 3600-9800 Å wavelength range with a resolution of about 20 Å (slit width 1.5″). The total exposure times on the two epochs were 15400 and 7600 sec, respectively.

The first attempt to recover SN 1986E was made on May 16, 1993 (phase 7.1 yr) using the 3.6m telescope with EFOSC. Two 3600 sec exposures were obtained using the B300 grism and the Tek512 CCD #26. This combination gives a wavelength range 3500-7000 Å and, using the 1.5″ wide slit, a resolution of 20 Å. The seeing was $FWHM = 1.4″$. The resulting spectrum was quite noisy but a broad H$\alpha$ emission was unambiguously present.

SN 1986E was observed again on March 16, 1994 (7.9 yr) with the New Technology Telescope (NTT) and the ESO Multi-Mode Instrument (EMMI). Three 3600 sec exposures were obtained using the grating #2 which in combination with the Tek2048 CCD #36 give a wavelength range 4500-9000 Å. The slit was 1.5″ wide and the resolution about 12 Å. Fair weather conditions (seeing $FWHM = 1.1″$) and the good performance of the new F/5.2 EMMI camera were crucial for the improvement of these observations compared with those of 1993.

Spectra were reduced using the IRAF packages. Wavelength calibration was obtained from the He-Ar comparison lamp and flux calibration by observations of standard stars from the list of Hamuy et al. (1992).

In the case of SN 1986E, to help positioning the slit, both at the 3.6m and at the NTT, R band exposures were obtained but, because of the presence of the nearby star, exposure times were limited to 30 sec. On these frames the SN is barely visible and the measure of the SN magnitude difficult. Estimates, obtained using the ROMAFOT package in MIDAS, are consistent with the SN magnitude being constant at the two epochs at $R = 22 \pm 0.5$.

## 3 SN 1986E

The spectra of SN 1986E at the two epochs are shown in Fig. 1. Allowing for the poor S/N ratio of the 7.1 yr spectrum, there is no evidence of evolution either in the flux or in the relative line intensities between the two epochs. The spectra are dominated by two broad emission lines centered at about 6300 and 6580 Å. These lines are readily identified with H$\alpha$ and [O I] $\lambda\lambda$6300,6364. Both were already strong in the spectrum at 9-10 months (Filippenko & Shields 1986; Schlegel et al. 1995) and are also the dominant features in the spectrum of the other recovered SNII, e.g. SN 1979C (Fesen & Matonick 1993) and SN 1980K (Fesen & Matonik 1994). The 7.9 yr spectrum extends up to 9000 Å and shows another broad feature between 7100 and 7400 Å, with a main peak at 7280 (7253)[†] Å and the hint of a second peak at about 7150 (7123) Å. A similar structure is present also on the late spectrum of SN 1980K, measured at 7100 and 7280 Å (Fesen & Matonik 1994). It is probable that the red component is due to [Ca II] $\lambda\lambda$7291,7324, which is a strong line in the nebular phase of SNII. The alternative identification with [O II] $\lambda\lambda$7320,7330 seems less probable because of the high excitation energy and small collisional strength of this transition compared with [O I]$\lambda\lambda$6300,6364 and [O III] $\lambda\lambda$4959,5007 (Chevalier & Fransson 1994). The identification of the blue component is more difficult: following Fesen & Matonick (1994) we indicate [Fe II] $\lambda$7155 as the most likely explanation for this feature. Note that other [Fe II] lines, in particular [Fe II] $\lambda$5159 and $\lambda$5273, may also be present.

Clearly visible are two narrow lines centered at 4978 (4959) and 5028 (5009) Å which are due to [O III] $\lambda\lambda$4959,5007 and appear unresolved on our spectrum (FWHM= 700 km s$^{-1}$). This fact, the close correspondence with the expected rest frame position and the relative intensity ratio suggests that they arise either from an underlying HII region or from circumstellar material surrounding the SN (the same origin can have the [N II] lines which we will mention later on). A small contribution from a broad [O III] component may also be present. In any case, the relative strength of this feature is much less than the corresponding one in SN 1979C or SN 1980K. There may also be an excess of flux in the region 5700-5900 Å which could be attributed to Na ID doublet or to He I $\lambda$5876 (flux $\leq$ 0.3 10$^{-15}$erg cm$^{-2}$ s$^{-1}$).

[†] In parenthesis is the wavelength in the galaxy rest frame adopting 1118 km s$^{-1}$ for the recession velocity of NGC4302 (Tully 1988)

**Table 1.** Comparison of the H$\alpha$ line widths of SN 1986E with those of other recovered SNII

| SN | age [yr] | centroid* [Å] | ZI velocities [km s$^{-1}$] |
|---|---|---|---|
| 1986E | 7.1 | 6520 | $-7000$ to $+2500$ |
|  | 7.9 | 6545 | $-6000$ to $+2600$ |
| 1970G$^a$ | 22 | 6558 | $-5400$ to $+5300$ |
| 1979C$^b$ | 11-12 | / | $-6280$ to $+6290$ |
| 1980K$^c$ | 7.8 | 6551 | $-6100$ to $+6300$ |
|  | 11.9 | 6561 | $-5400$ to $+5350$ |

\* Wavelengths are in the parent galaxy rest frame
Data from: *a*) Fesen (1993), *b*) Fesen & Matonick (1993), *c*) Fesen & Matonick (1994))

In the spectrum there is no clear evidence for other lines, and in particular H$\beta$ is not visible. The observed Balmer decrement, H$\alpha$/H$\beta \geq 10$ in the broad component, is much larger than the predicted for case B.

The $FWHM$ of the H$\alpha$ line corresponds to 4000 km s$^{-1}$ which is much larger than the line width measured 9–10 months after explosion, $FWHM = 2500-3000$ km s$^{-1}$ (Filippenko & Shields 1986; Schlegel et al. 1995). The centroid and the zero intensity (ZI) velocities of the H$\alpha$ emission are reported in Tab. 1 and compared with those of other SNII. In the 7.9 yr spectrum, the H$\alpha$ emission extends from $-6000$ to $+2600$ km s$^{-1}$ with a peak at zero velocity, whereas at 7.1 yr the line had a similar shape but larger expansion velocities, $-7000$ to $+2500$ km s$^{-1}$. The [O I] $\lambda\lambda$6300,6364 doublet is clearly displaced to the blue compared with the rest frame position: the centroid of the line in the 7.9 yr spectrum is measured at 6305 (6281) Å. The $ZI$ velocity of the [O I] emission is impossible to measure because its double structure is blended with the extended blue wing of H$\alpha$. From the blue wing we measure $-7350$ km s$^{-1}$. Although the line is very broad and noisy, also the [Ca II] $\lambda\lambda$7291,7324 feature is clearly blue-shifted at about 7280 (7253) Å.

As better seen in Fig. 2, the H$\alpha$ profile is asymmetric with a steep red wing. A similar H$\alpha$ profile was seen in SN1988H about 14 months past the explosion (Turatto et al. 1993) and it was attributed to dust formation in the ejecta which absorbs a significantly greater fraction of the line emission from the far–side, receding material. Assuming that the blue wing is unaffected by the dust, the flux deficiency in SN 1986E is $\sim 35\%$. This gives further support to the growing evidence of dust in SN ejecta at different stages of evolution: in particular SN1957D 31 yr past maximum (see Sect. 4), SN1979C at 12 yr (Fesen & Matonick 1993), SN1987A at 1.5 yr (Danziger et al. 1991) and SN 1988H at 1.2 yr (Turatto et al. 1993).

In the 7.9 yr spectrum there is also evidence of structure in the H$\alpha$ profile (Fig. 2): two peaks are displaced symmetrically by about 5000 km s$^{-1}$ carrying 10–15% of the total H$\alpha$ emission (two other peaks displaced by about 1000 km s$^{-1}$ are in close correspondence with the [N II] $\lambda\lambda$6548,6583 lines and therefore may be related to an underlying HII region or to circumstellar material).

In Tab. 2 we compare the line luminosities of SN 1986E at late epoch with those of other SNII. For consistency, the line luminosities are corrected only for the galactic extinction. The absolute line luminosities have large uncertainties because: a) the objects are very faint and the absolute flux calibration may introduce errors up to 50%; b) the distances of the parent galaxies are known only with a 30% accuracy; c) the absorption in the parent galaxy is, in most cases, unknown. Instead, relative line fluxes are much more reliable (±15%), because they do not depend on the distance and only slightly on reddening.

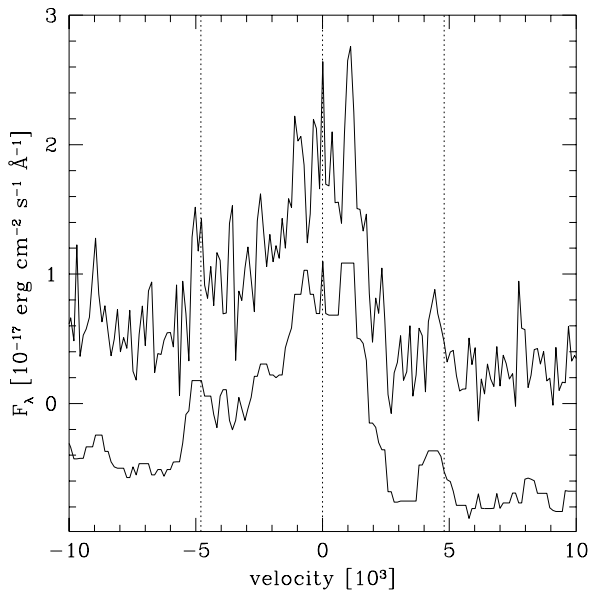

**Figure 2.** The Hα profile at 7.9 yr in velocity coordinates. The lower line shows a mild filtering of the original spectrum. As a guide for the eye vertical lines are drawn at -4800, 0, +4800 km s$^{-1}$

The comparison shows that the late Hα luminosity of SN 1986E is very similar to that of SN 1979C and one order of magnitude brighter than in SNe 1970G and 1980K. With regard to the relative line ratios the [Ca II]/Hα and [O I]/Hα are very close in all these SNe while the [O III]/Hα is significantly smaller in SN 1986E (by a factor 3 to 5).

## 4  SN 1957D

We now compare the observations of SN 1986E with those of SN 1957D which is, at present, the oldest example of a spectroscopically recovered SNe. Unfortunately, SN 1957D was not observed near maximum light and therefore spectroscopic and/or photometric classification is not available. This SN was recovered as a non-thermal radio source 25 yr after the explosion by Cowan & Branch (1985) and some years later, guided by the radio position, optical spectra were obtained (Long et al. 1989; Turatto et al. 1989). The spectra obtained at ESO 31.3 yr and 33.1 yr after outburst are shown in Fig. 4. On both epochs the dominant feature is [O III] λλ4959,5007; after deblending the line width was found $FWHM = 2500$ km s$^{-1}$, which allows the association of the emission with the expanding debris of the SN explosion.

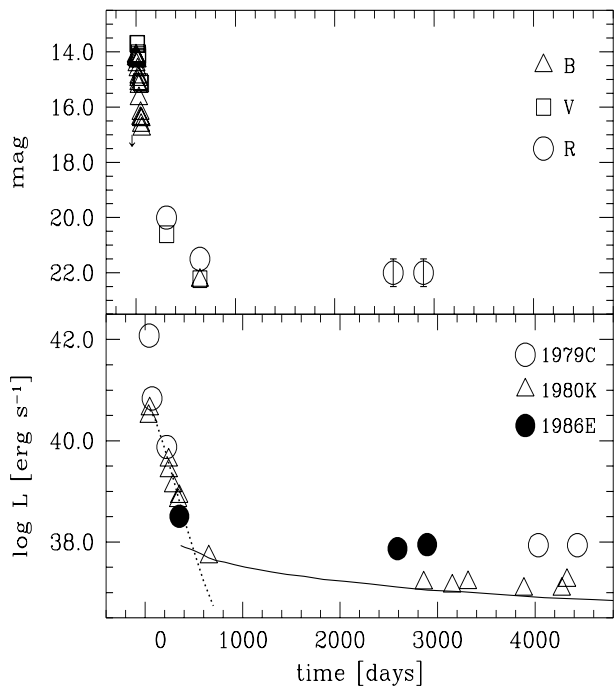

**Figure 3.** *Top panel*: BVR light curve of SN 1986E. The measurements at 308 days are derived from the spectrum of Schlegel et al. (1995). *Bottom panel*: Hα luminosity evolution of SN 1986E compared with those of SNe 1979C and 1980K. The dotted line gives the predicted Hα luminosities in the radioactive phase according to Chugai (1991) (model with $M_{ejecta} = 5$ M$_\odot$ and $M_{Ni} = 0.075$ M$_\odot$) whereas the solid line is the luminosities predicted in the circumstellar interaction model of Chevalier & Fransson (1994) ("power law" model).

Surprisingly there has been significant luminosity evolution between the two epochs with almost a factor 3 decrease of the [O III] emission in 2 yr. This is in agreement with the report by Long et al. (1992) (the value reported in the preliminary description of the 31.3 yr spectrum by Turatto et al. 1989 was incorrect owing to a software bug in the reduction package).

Because of the good S/N ratio, we could also detect broad emission of [O I] λλ6300,6364 with a FWHM identical, within the uncertainties, to that of the [O III] emission. Also [O I] shows a significant luminosity decline and, as a consequence, the line ratio [O III]/[O I]∼5 remains constant,within the uncertainties. Finally both of these features appear blue-shifted compared with the rest frame position by about 600 km s$^{-1}$. We have already interpreted this as being due to dust formation in the ejecta (Turatto et al. 1989).

In the spectra the unresolved lines of Hα+[N II], Hβ and, only in the better S/N spectrum of April 1989, of [S II] λλ6717,6731 are also visible. All these emissions are centered at the rest wavelengths and show no flux evolution and therefore, as also suggested by Long et al. (1992), they probably arise from the underlying HII region.

Because there is no evidence in the spectra of broad Hα emission ([O III]/Hα > 5.3) one might be tempted to con-

**Table 2.** Comparison of the late line luminosity of SNII

| SN | age [yr] | distance [Mpc]$^d$ | E(B-V)$^e$ | flux [$10^{-15}$ erg cm$^{-2}$ s$^{-1}$] H$\alpha$ | [Ca II] | [O I] | [O III] | H$\alpha$ Lum.$^f$ | [Ca II]/H$\alpha$ $g$ | [O I]/H$\alpha$ $g$ | [O III]/H$\alpha$ $g$ |
|---|---|---|---|---|---|---|---|---|---|---|---|
| 1986E | 7.9 | 16.8 | 0.02 | 2.5 | 1.1 | 2.2 | $\leq 0.7$ | 8.9 | 0.4 | 0.9 | $\leq 0.3$ |
| 1957D | 31.3 | 4.7 | 0.14 | < 0.9 | $\leq 0.6$ | 0.9 | 4.8 | < 0.3 | | > 1.0 | > 5.3 |
| " | 33.1 | " | " | < 0.7 | $\leq 0.2$ | 0.4 | 1.7 | < 0.3 | | > 0.6 | > 2.5 |
| 1970G$^a$ | 22 | 5.4 | 0.00 | 1.8 | | 1.0 | $\leq 1.0$ | 0.6 | | 0.7 | $\leq 0.7$ |
| 1979C$^b$ | 11–12 | 16.8 | 0.01 | 2.5 | 1.6 | 2.5 | 3.5 | 8.8 | 0.6 | 1.0 | 1.4 |
| 1980K$^c$ | 8–12 | 5.5 | 0.40 | 1.7 | 0.7 | 1.3 | 1.1 | 1.6 | 0.4 | 0.8 | 0.9 |

Data from: *a*) Fesen (1993); *b*) Fesen & Matonick (1993); *c*) Fesen & Matonick (1994);
*d*) Tully (1988);
*e*) only galactic contribution from Burstein & Heiles (1982);
*f*) units $10^{37}$ erg s$^{-1}$;
*g*) the error on the relative intensity ±15%.

clude that hydrogen is absent and that the ejecta is oxygen-rich. In this case SN 1957D could have been a type Ib/c SN. Alternatively, as suggested by Chevalier & Fransson (1994), the dominance of the [O III] emission may be due to the particular physical conditions of the emitting gas and SN 1957D could have been a normal type II SN, similar to SN 1986E.

## 5 DISCUSSION

One of the most striking properties of SN 1986E is that, as shown in Fig. 3, the R luminosity at 8 yr was almost the same as it was 2 yr after the explosion. The behaviour of the R magnitudes may be reflecting the behaviour of the strong [O I] and H$\alpha$ lines which lie within the bandwidth of this filter. In the bottom panel of Fig. 3 we compare the H$\alpha$ luminosity evolution of SN 1986E with that of SN 1979C and SN 1980K. The three SNe exhibit a very similar behaviour: a rapid luminosity decline in the first 2 yr is followed by a long phase (> 10 yr) where the luminosity is almost constant. The early decline is explained assuming that the input energy derives from the radioactive decay of $^{56}$Co. In particular, the simple radioactive model of Chugai (1991) adopting an initial $M_{Ni} = 0.075$ M$_\odot$ and a $M_{ejecta} = 5$ M$_\odot$ give a very good fit of the first two years of observations (dotted line in Fig. 3).

However, the radioactive decay of $^{56}$Co can give only a marginal energy contribution at the latest phase and an additional energy input must be sought. Because the emission line widths are larger than those measured in the radioactive epochs the emitting regions have to be located in the outer ejecta material. It is now believed that this can be due to kinetic energy released by the interaction of the ejecta with the circumstellar material, whereas other energy sources such as other radioactive species or a pulsar seem less likely (Chugai 1990; Chugai 1991; Chugai & Danziger 1994; Fesen & Matonik 1994; Chevalier & Fransson 1994).

Chevalier & Fransson (1994) have proposed a model of interaction of the SN ejecta with the progenitor wind in which the hard radiation from the forward and reverse shock fronts heats both the material in the outer ejecta and that in the shell between the two shocks. The model predicts that strong optical emission lines originate from these two regions and the expected spectrum is found qualitatively consistent with the observed spectra of SNe 1979C and 1980K. The circumstellar interaction depends on the density profile, velocity and geometry of the expanding ejecta and circumstellar wind. Chevalier & Fransson (1994) have shown that a good fit with the late observations of SN 1980K (solid line in Fig. 3 can be obtained adopting a power law density profile (with $n = 8$) for the outer ejecta and $\dot{M}_{-5} v_{w1}^{-1} = 5$ where $\dot{M}_{-5}$ is the mass loss rate in units of $10^{-5}$ M$_\odot$ yr$^{-1}$ and v$_{w1}$ is the presupernova wind velocity in units of 10 km s$^{-1}$ ). Because the absolute luminosity depends on the mass loss rate, the higher H$\alpha$ luminosities of SNe 1979C and 1986E compared with that of SN 1980K at similar epochs indicate a higher mass loss rate in the former SNe.

We can try to relate the late optical emission of Linear SNII to the presupernova history of these objects. Usually, the differences in the early light curves of Linear and Plateau SNII are related to the progenitor envelope masses at the time of explosions: these are significantly smaller in Linear SNII, possibly by a factor 5 (Swartz et al. 1991; Cappellaro et al. 1994), because of much stronger mass loss during progenitor evolution. Linear SNII are strong radio emitters (Weiler et al. 1989) and a clear correlation between radio luminosity, estimated mass loss and late–time H$\alpha$ luminosity has been found by Fesen (1993). In particular for SN 1979C, which shows a high H$\alpha$ luminosity, the estimated mass loss rate is a factor 3 to 5 times higher than in 1970G and 1980K. Because of its high H$\alpha$ luminosity we suggest that SN 1986E is now a powerful radio source with observable flux at 20cm in the range 0.5 − 5 mJy, the higher value being favored by the similarity with SN 1979C. We are not aware of attempts to detect SN 1986E: radio observations of this SN are therefore encouraged.

The observed [O I]/H$\alpha$ ratio in SN 1986E ([O I]/H$\alpha$ = 0.9) is similar to that of other SNII observed at late epochs (the 'power law' model with $n = 8$ of Chevalier & Fransson 1994 predicts [O I]/H$\alpha$ = 0.45). Instead [O III]/H$\alpha$ < 0.3 is significantly smaller in SN 1986E than in other objects. This may be related to the *younger* age of SN 1986E which at the time of observations was only 7-8 years. The models by Chevalier & Fransson (1994) predict an abrupt increase of the [O III] emission starting about ten years after explosion. Indeed, according to this model, in 20 years the spectrum of SN 1986E is expected to become [O III] dominated, [O III]/H$\alpha$=6, consistent with the observations of SN 1957D, [O III]/H$\alpha$ > 5.3. It must be noted, however, that the predicted value at 8 yr, [O III]/H$\alpha$=1.5, is a factor 5 higher than the observed one in SN 1986E. Moreover, in the coeval spectrum of SN 1980K (7.8 yr) the [O III] emission was already developed and no significant evolution of the line ratios was seen in later spectra (Fesen & Matonik 1994). Chevalier & Fransson (1994) point out that the [O III]/H$\alpha$ ratio is sensitive to the exact value of the reverse shock velocity, as well as the ejecta density. From this point of view the smaller [O III]/H$\alpha$ ratio in SN 1986E could result from a flatter density profile than in SN 1980K.

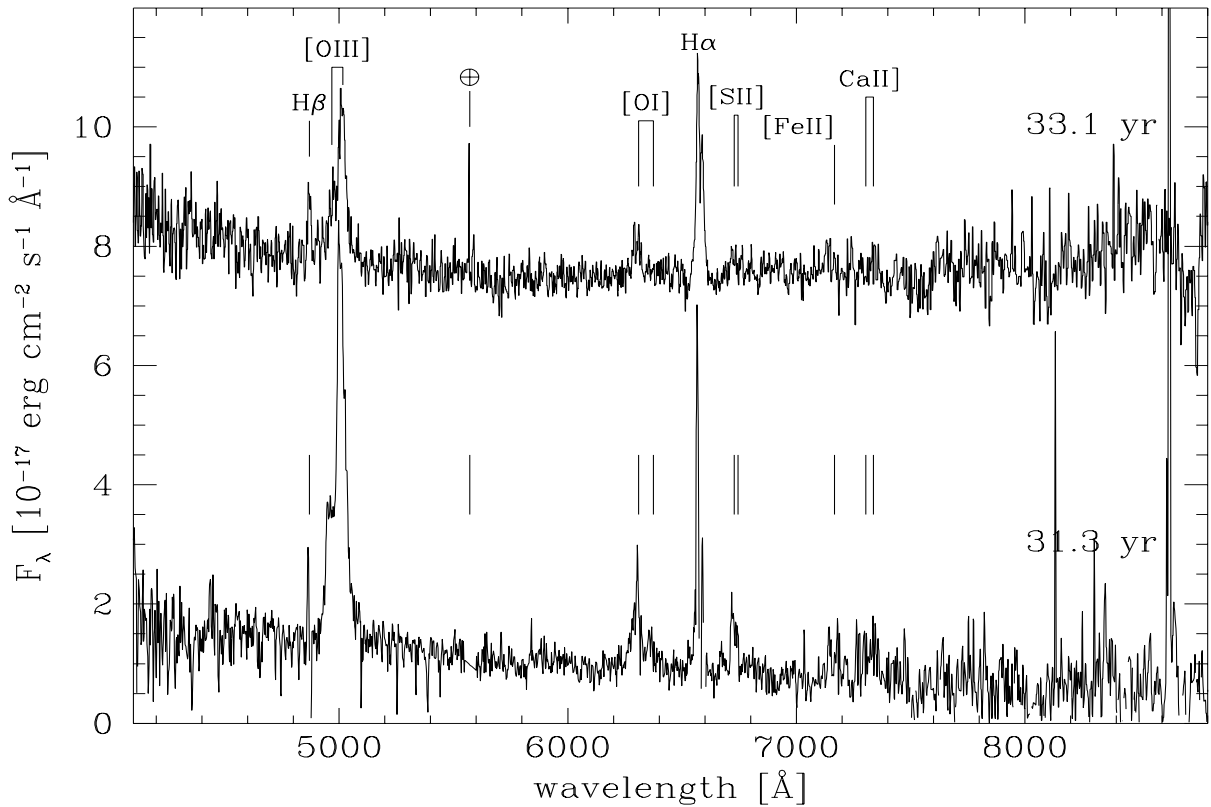

**Figure 4.** Spectra of SN1957D 31.3 yr and 33.1 yr after explosion. The latter has been shifted in the vertical axis by 8 units. Vertical lines mark the positions of the lines in the galaxy rest frame ($v_{hel} = 518$ km s$^{-1}$).

One of the problems of the Chevalier & Fransson (1994) model is that the H$\alpha$ luminosity is predicted to decline on a time scale of a few years. This was not observed in SN 1980K which showed an almost constant flux and line ratio in the period 7.8 yr to 11.9 yr and in SN 1986E which declined only a factor $\sim$ 4 in the period 1 to 8 yr instead of a factor 10 as predicted. However, it is clear that the light curve of this kind of object depends on the mass loss history of the progenitor. The best example is given by the sharp luminosity decline exhibited in the period 1989-1991 by SN 1957D, much larger than that predicted by the particular Chevalier & Fransson model which adopts a uniform wind distribution but which may be explained if there is a density drop of the presupernova wind at a radius of about $\sim 10^{18}$ cm (Long et al. 1992).

Another characteristic feature of SN 1986E is the H$\alpha$ profile with a sharp red wing and, in the 7.9 yr spectrum, two symmetric satellite emissions displaced by $\pm 5000$ km s$^{-1}$, falling close to the edge of the main emission profile. Nothing similar appears in the [O I] line, but the S/N is poor for this line. The other SNII showed different features, e.g. twelve years past maximum SN 1979C had also an asymmetric H$\alpha$ profile but showed a sharp blue edge (Fesen & Matonick 1993) whereas SN 1980K showed a flat–topped H$\alpha$. Recently, a two-peaked structure of the H$\alpha$ emission was seen in the hydrogen lines of the type IIb SN 1993J 500 days after explosion and was interpreted as originating in a non-spherical emission region (Spyromilio & Leibundgut 1995).

The two satellite emission lines flanking H$\alpha$ in the spectrum of SN 1986E could originate either from a non-spherically shaped (or even jet–like) ejecta aligned somewhere near the line of sight or from a ring. Whereas a non-spheroidal ejecta could result because of asymmetries in the explosion, a cool ring could form in the region between the forward and reverse shocks after the collision of the ejecta with the circumstellar material if the progenitor wind had an enhanced density in the equatorial plane. In both cases the bulk of the observed Hα emission should originate from a spherically symmetric interaction of the ejecta with the circumstellar material. We note that because the satellite emissions are close to the Hα edge velocity the equatorial plane must be close to the line of sight. The two scenarios cannot be tested on the basis of the available material and, because the presence of these kinds of structures have strong implications for the models, we stress the necessity to obtain higher signal-to-noise spectra of this SN.

## 6 CONCLUSIONS

SN 1986E is the fourth Linear SNII, after SNe 1970G, 1979C and 1980K, recovered in the optical several years after explosion. The objects share many similar characteristics namely

(i) the Hα luminosities remain almost constant starting 2 years after explosion and for at least 10-15 years;
(ii) the spectrum is dominated by the Hα line with major contributions from [O I] $\lambda\lambda 6300,6364$ and [Ca II] $\lambda\lambda 7291,7324$ and in two SNe, 1979C and 1980K, from [O III] $\lambda\lambda 4959,5007$;
(iii) SN 1970G, 1979C and 1980K show strong radio emission (radio observations of SN 1986E were probably never attempted).

We believe that these emissions result from the interaction of the ejecta with the circumstellar material. The circumstellar interaction model of Chevalier & Fransson (1994) is in fair agreement with the observations of SN 1980K when adopting a power law density profile of the ejecta outer regions. The fit of the observations of SN 1986E would requires a higher mass loss rate and flatter density profile compared with SN 1980K. If there had been an intermediate velocity component due to clumpiness in the CSM as in the model of SN1988Z proposed by Chugai & Danziger (1994), it is likely that it has, after 8 years, faded below the resolution limit of our spectra. Apart from this fading, one would need a significantly better S/N ratio to detect it.

SN 1986E is a good candidate for testing the circumstellar interaction models of Chevalier & Fransson. In the next decade, we expect an increase of the [O III] emission relative to Hα. Eventually the spectrum should become similar to that observed of SN 1957D 30 yr after explosion that probably is a normal SNII with a dense circumstellar medium and no need of unusual oxygen abundance.

Because the luminosity of the SN at these phases strongly depends on the density of the circumstellar wind, and on density and velocity profile of the SN, continuing observations of SN 1986E and SN 1957D can be used to reconstruct the structure and mass loss histories of the progenitor stars.